\documentclass[twocolumn,groupedaddress,nofootinbib,showpacs]{revtex4}

\usepackage{graphicx}
\usepackage{dcolumn}

\usepackage{bm}
\usepackage{color}
\input{epsf}
\begin{document}

\title{Static strings in global AdS space and quark anti-quark potential}

\author{C. A. Ballon Bayona}
\email{ballon@if.ufrj.br}
\affiliation{Instituto de F\'{\i}sica, 
Universidade Federal do Rio de Janeiro, Caixa Postal 68528, 
RJ 21941-972 -- Brazil}
\author{Cristine N. Ferreira}
\email{crisnfer@pq.cnpq.br}
\affiliation{ N\'ucleo de Estudos em F\'isica, Centro Federal de Educa\c c\~ao Tecnol\'ogica de Campos,
Campos dos Goytacazes, RJ 28030-130,  Brazil}

\begin{abstract}  
We investigate the finite temperature quark anti-quark problem in a compact space $S^{n-1}\times S^1$ by considering static strings in global $AdS_{n+1}$ space with $n\ge 3$. For high temperatures we work in the black hole metric where two possible solutions show up : the big black hole and the small black hole.  Using the AdS/CFT correspondence, we calculate the quark anti-quark potential (free energy) as a function of the distance. We show that this potential can be intrepeted as confining for the $AdS$ space and deconfining for the big black hole. We find for the small black hole a confining limit for the potential but this solution is instable following the Hawking-Page criteria. Our results for the free energy   reinforce the Witten interpretation of the confinement/deconfinement transition as the dual of the well-known Hawking-Page transition.
\end{abstract}

\pacs{ 11.25.Tq ; 11.25.Wx ; 12.38.Aw }

\maketitle

\section{Introduction}

Confinement of quarks is a very old problem that until now has not been solved. At low energies perturbative QCD can not be applied because the coupling constant is large. Effective theories and lattice QCD have given important contributions to the understanding of this problem. A very useful approach to the confinement problem is the study of large N gauge theories because of the possibility of obtaining analytical and numerical results for large (`t Hooft) coupling in a simple way via the AdS/CFT correspondence \cite{Maldacena:1997re,Aharony:1999ti}.
This correspondence relates large $N$ gauge theories on a $n$ dimensional manifold $\mathcal{M}$ to 10-d string theories on $AdS_{n+1}\times W$ being $\mathcal{M}$ the boundary of $AdS$ and $W$ some compact space. In particular, in the context of AdS/CFT correspondence Witten  showed that ${\cal N}=4$ Super Yang-Mills (SYM)  theory  in $S^{3} \times S^{1}$ at strong coupling experiments a confinement/deconfinement transition that corresponds holographically to a gravitational transition between  $AdS_5$ and black hole $AdS_5$ defined in global coordinates \cite{Witten:1998qj,Witten:1998zw}. This gravitational transition was studied before by Hawking and Page \cite{Hawking:1982dh}. The confinement criteria stated in \cite{Witten:1998qj,Witten:1998zw} is the behavior of the free energy of large N $\it{N}=4$ SYM theory in $S^{3}\times S^1$. The theory is confining at low temperatures because the free energy is of order 1 (color singlet contribution) while is deconfining at high temperatures because the free energy is or order $N^2$ (gluons contribution). This criteria can be extended to the cases $S^{n-1}\times S^1$.   

Finite temperature QCD lives in $R^{3}\times S^1$ which is the boundary of $AdS_5$ in Poincar\'e coordinates. SYM theory in $R^{3}\times S^1$ is not confining as can be seen using the Wilson loop prescription \cite{MaldaPRL}.  This prescription relates a Wilson loop of the gauge theory to the world-sheet area of a static string living in $AdS_5\times S^5$ in Poincar\'e coordinates. In particular, the free energy of a quark anti-quark pair in $R^{3}\times S^1$ can be calculated from the correlator of two Polyakov loops and exhibits a coulombian behavior. This behavior can be interpreted as deconfining because the free energy vanishes when the quark anti-quark distance goes to infinity. The Wilson loop prescription was generalized to other curved spaces and a confinement criteria was established \cite{Kinar:1998vq}. The quark anti-quark free energy at high temperatures for the space $R^{3}\times S^1$ was calculated in \cite{Rey:1998bq,Brandhuber:1998bs,BoschiFilho:2006pe} by considering an $AdS_5$ black hole in Poincar\'e coordinates. Other Wilson loop calculations involving gauge/string duality can be found at \cite{RY,Greensite:1998bp,Bigazzi:2004ze}. 

There are some phenomenological models that introduce confinement in $R^{3}\times S^1$. A well succeeded one is the hard wall model which consists on introducing a hard wall on Poincar\'e $AdS_5$ space interpreted holographically as an infrared cut-off for the gauge theory. This model was motivated on the calculation of  scattering amplitudes \cite{Polchinski:2001tt} and leads to confinement at low temperatures \cite{BoschiFilho:2005mw}. At high temperatures there is a confinement/deconfinement transition corresponding to a Hawking-Page transition between (Poincar\'e) $AdS$ and (Poincar\'e) black hole $AdS$ space \cite{Herzog:2006ra,BallonBayona:2007vp}. Other calculations of quark anti-quark potential using phenomenological models can be found at \cite{Andreev:2006nw,Dorn:1998tu,Hartnoll:2006hr}.  

In spite of being a little far from real world, large $N$ gauge theories on $S^{n-1}\times S^{1}$ are physically interesting. For instance, it is possible to use perturbation theory in the small radius regime so that thermal transitions can be studied and connect them with the Hawking-Page and Hagedorn transitions which are important in string theory and quantum gravity \cite{Aharony:2005bq,Aharony:2003sx}.  
The mean of this work is to calculate the free energy of  a quark anti-quark pair living in  $S^{n-1}\times S^{1}$ space where $S^{1}$ represents the imaginary time coordinate with a period $\beta$ identified  with the inverse of the temperature. For this purpose we consider a Wilson loop containing two temporal lines and two spatial lines. The quark anti-quark free energy can be extracted from the correlator of two Polyakov loops. In order to relate our Wilson loop to the Polyakov loop correlator, the size of the temporal lines must be $\beta$.  The dual description consists in a static string with the quark and anti-quark at the end points. This string configuration gives the dominant contribution to the connected part of the Polyakov loop correlator (see for example \cite{Witten:1998zw,Rey:1998bq,Brandhuber:1998bs,Andreev:2006nw,Bak:2007fk}). Then what we mean by potential in this article is the quark anti-quark free energy obtained from the connected part of the Polyakov loop correlator. We follow the procedure of \cite{Kinar:1998vq} to calculate this potential. The space $S^{n-1}\times S^{1}$ is the  boundary of two spaces:  global $AdS$ and black hole $AdS$ in $n+1$ dimensions.  We first calculate explicitly the $n =3$ case where asymptotic limits for the potential are studied and then we generalize our results to the cases $n \ge 4$. We find a coulomb-like behavior of the free energy in the case of global $AdS_{n+1}$ and show that this behavior can be interpreted as confining. At high temperatures we calculate the quark anti-quark free energy for the two possible black hole solutions : the big black hole and small black hole in $AdS_{n+1}$. In the case $n=4$ our results for the big black hole agree with those obtained before by Landsteiner and Lopez \cite{Landsteiner:1999up} where it was defined a screening length for the quark anti-quark free energy. Here, we calculate this screening length as a function of the temperature and the horizon position. For the big black hole the screening length never reaches the maximal distance while for the small black hole this could happen. This will imply (for $n \ge 3$) that the big black hole $AdS_{n+1}$ is indeed deconfining at any (high) temperature while the small black hole could give confinement in the limit $T \to \infty$. The Hawking-Page criteria states that the small black hole solution is instable, so deconfinement is guaranteed at high temperatures. Our results indicate a confinement/deconfinement transition for the quark anti-quark free energy in $S^{n-1}\times S^{1}$ corresponding to a Hawking-Page transition from  $AdS_{n+1}$ to black hole $AdS_{n+1}$.
 
\section{Wilson Loops on the AdS boundary}

In this section let us calculate the quark anti-quark free energy on the compact space $S^{2}\times S^{1}$. This space is the boundary of the $AdS_4$ space in global coordinates. The metric of global $AdS_4$ with Euclidean signature is  

\begin{equation}
ds^2 = \Big( 1 + {r^2 \over R^2} \Big) d t^2 + {d r^2 \over \Big( 1 + {r^2 \over R^2} \Big)} + 
r^2\Big( d \theta^2 + \sin^2\theta d\varphi ^2\Big)\label{metricaadsrestrita}
\end{equation}
\noindent where $0\le \theta < \pi $ ,  $-\pi \le \varphi <\pi $ and $0<r<\infty$ . The time coordinate is compact : $0 \le t \le \beta$ with $\beta=1/T$ .

For simplicity we consider   $\theta = \pi/2$ and define a new  coordinate $x = R \varphi$ with dimension of length. The metric then
reads

\begin{equation}
ds^2 = \Big( 1 + {r^2 \over R^2} \Big) d t^2 + {d r^2 \over \Big( 1 + {r^2 \over R^2}\Big) } + 
{r^2\over R^2}dx ^2\label{metricaadsx}
\end{equation}

\begin{figure}
\centering
\includegraphics[width=5cm]{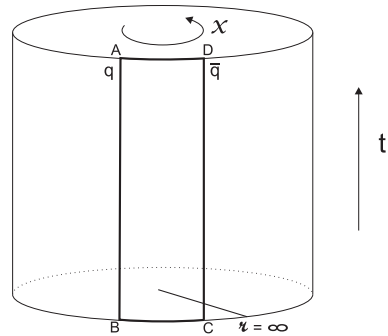}
\parbox{5in}{\caption{The Wilson loop on the boundary of $AdS_4$ \\with $\theta=\pi/2$. $AB$ and $CD$ are the corresponding \\Polyakov loops. The top and bottom bases of the solid \\ cylinder are identified. 
 }\label{Fig1}}
\end{figure}

\noindent where $-\pi R \le x <\pi R$. We consider the rectangular Wilson loop of Fig \ref{Fig1} living in the $AdS$ boundary ($r = \infty $). The quark and anti-quark are localized at $x = {L \over 2}$ and $x = -{L \over 2}$. The distance between the quarks $L$ has a maximum value $L_{\rm max }=\pi R$ because of the compactness of the $x$ coordinate.

The dual configuration consists on a static string living on $AdS$ with the quarks at the end points. The Nambu-Goto action for this string is 
\begin{eqnarray}
S &=& {1 \over 2 \pi \alpha'} \int d \sigma d\tau \Big( \sqrt{det(G_{M N} \partial_{\tau} X^M \partial_{\sigma} X^N }\Big) \nonumber \\
&= & {\beta \over 2\pi \alpha'} \int_{-L/2}^{L/2} d x \sqrt{f(r)^2 + g(r)^2 (\partial_{x} r)^2 }
\end{eqnarray}
where

\begin{eqnarray}
f^2 &= g_{tt} g_{x  x} &= \Big(1+ {r^2 \over R^2}\Big) {r^2 \over R^2}  \nonumber \\
g^2 & = g_{tt} g_{r r} & = 1
\end{eqnarray}

 \noindent and we parameterized the string configuration by $\tau = t$ and $\sigma = x$  valid for $-\pi R/2 \le x \le \pi R/2 $ \,  (the other region should be parameterized in other way but it is not necessary to work there because of the symmetry of the problem).
 
The solutions for the static string configuration (corresponding to a minimum action) represent geodesics of the AdS space where the string reaches a minimum for the radial coordinate $r_0$ (see Fig 2). Using the prescription of \cite{Kinar:1998vq} we find 

\begin{eqnarray}
L & = & 2\int_{r_0}^{\infty}dr {g(r) \over f(r)} \Big[ {f^2(r_0) \over f^2(r) - f^2(r_0)}\Big]^{1/2} \nonumber\\  
& =&   \,  \int_{1}^{\infty}  \frac{2 R \sqrt{1+\Lambda^2}dy}{y^4\sqrt{\Lambda^2+\frac{1}{y^2}}\sqrt{1-\frac{1}{y^2}} \sqrt{\frac{1+\Lambda^2}{y^2}+\Lambda^2}}
\end{eqnarray}

\noindent where $\Lambda = {r_0 \over R}$ and $y = {r \over r_0}$.

\begin{figure}
\centering
\includegraphics[width=4.5cm]{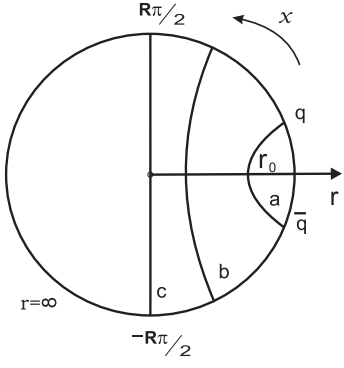}
\parbox{2in}{\caption{Schematic representation of geodesics 
in AdS space.}\label{Fig2}} 
\end{figure} 

The regulated free energy is given by 

\begin{figure}
\centering
\includegraphics[width=7.5cm]{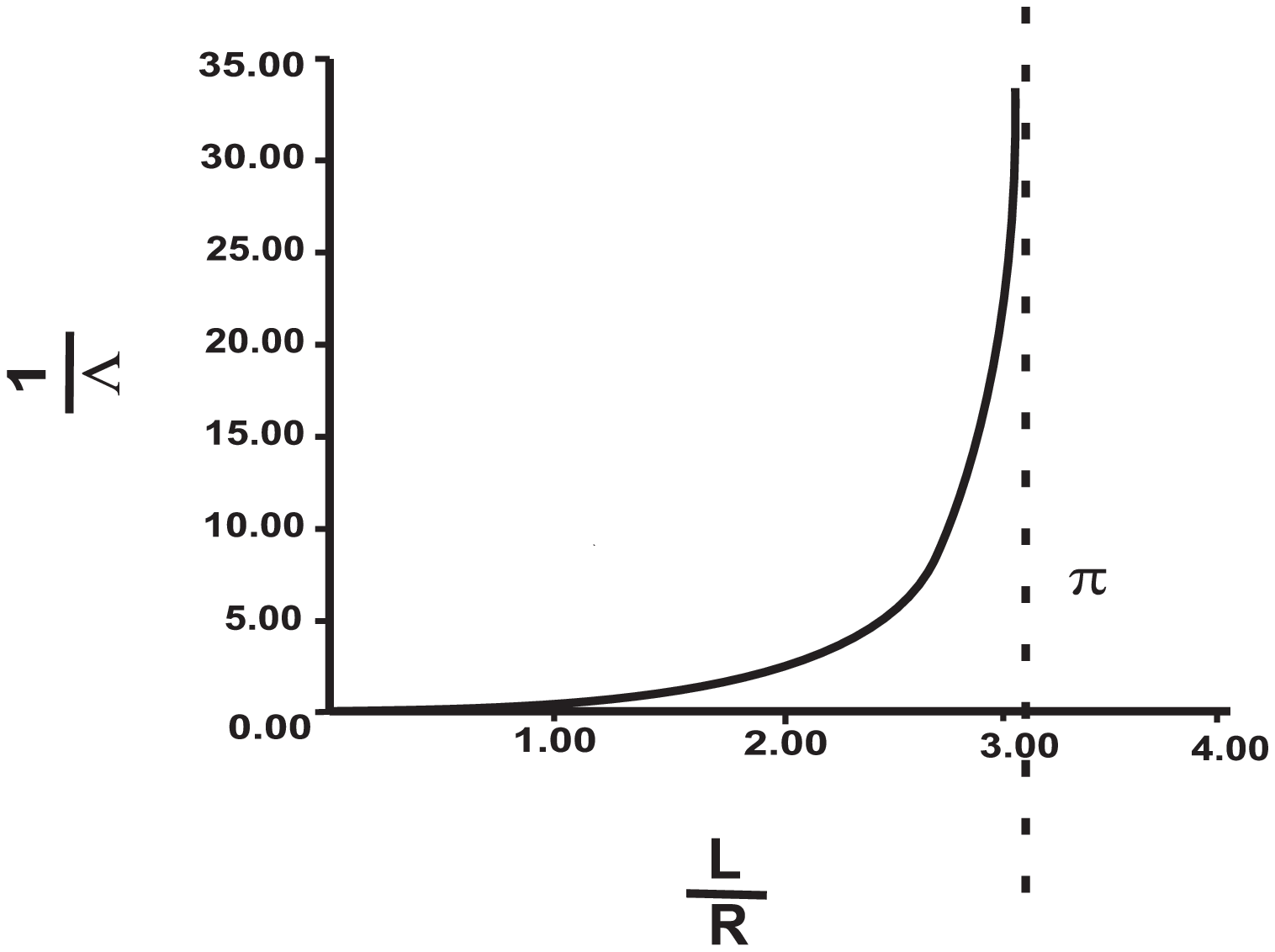}
\parbox{2in}{\caption{$\! \! \! $ ${1\over \Lambda }$ vr ${L \over R}$ for global$AdS_4$.}\label{Fig4}} 
\end{figure}

\begin{eqnarray}
F & \! \! \! \! \! \! \! \! = &\! \! \! \! \! \!\! \! \! \!\! \! \!{1 \over \pi \alpha'} \int_{r_0}^{\infty} \! \!dr \frac{g(r)f(r)}{\sqrt{f^2(r)-f^2(r_0)}}- {1 \over \pi \alpha'} \int_0^{\infty}\! \! \!  dr g(r) \nonumber\\
&= {\Lambda R \over \pi \alpha'} &  \! \! \!\! \! \!  \left[ \!  \int_{1}^{\infty} \! \! \! \! \!  dy \! \! \left (\! \! \frac{\sqrt{\frac{1}{y^2}+\Lambda^2}}{\sqrt{1-\frac{1}{y^2}}\sqrt{\Lambda^2+\frac{1+\Lambda^2}{y^2}}}-1 \! \! \right )-  1\right]
\end{eqnarray}

\noindent where the second integral of the first line represents a regulator interpreted as the sum of the quark anti-quark masses.

\begin{figure}
\centering
\includegraphics[width=7.5cm]{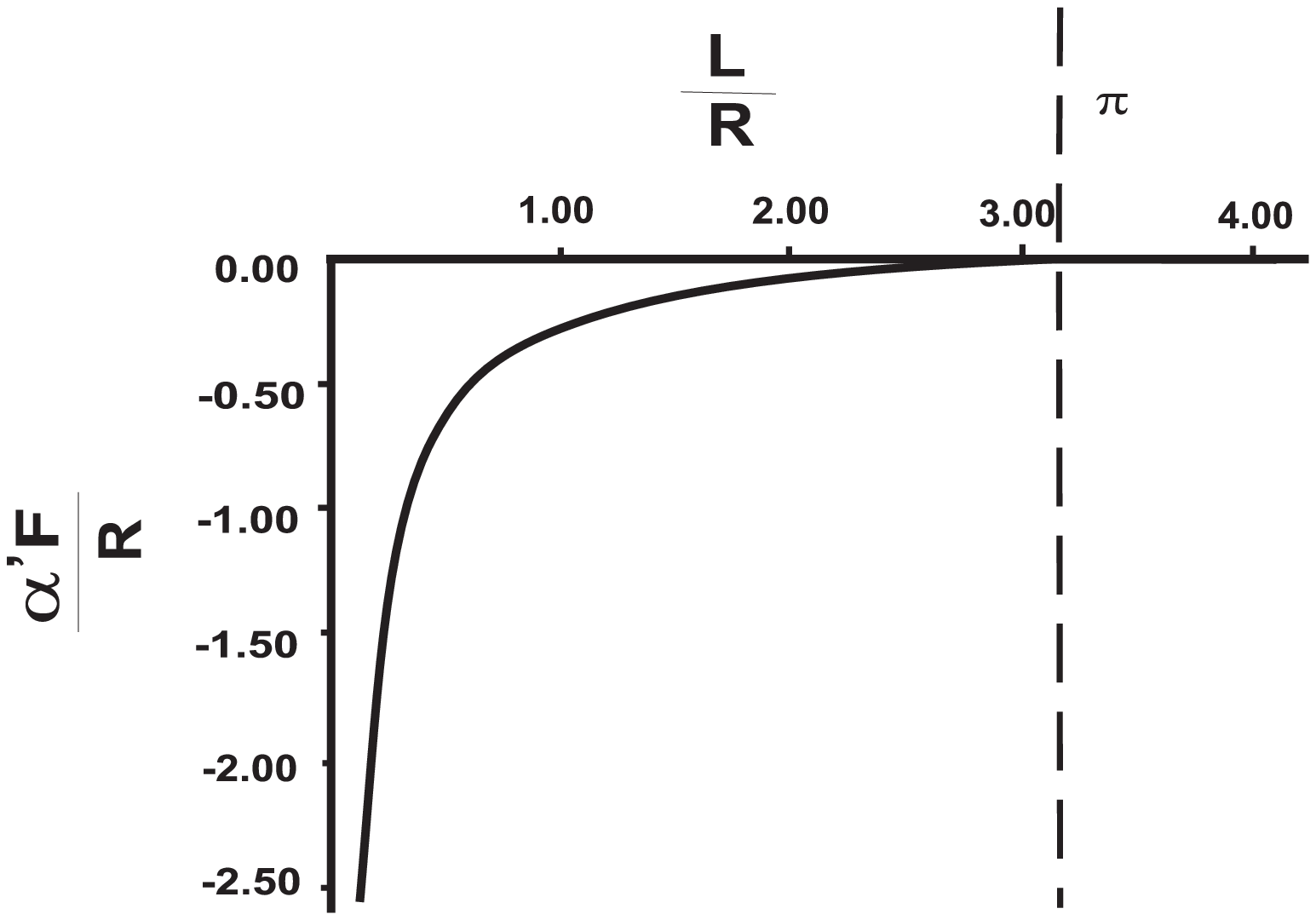}
\parbox{2in}{\caption{${ \alpha'F\over R}$  vs $L\over R$ for global$AdS_4$ . }\label{Fig5}}
\end{figure} 

The integrals for $L$ and $F$ are elliptical so we calculate them numerically. We find a free energy with coulombian behavior at small distances and that goes to zero only when $L$ goes to $L_{\rm max}$. We will explain in the last section why this behavior for the free energy can be interpreted as confining. Our results are shown in Figs. \ref{Fig4} and \ref{Fig5}.   

It is interesting to analyze the behavior of the free energy in the asymptotic limits $L<<R$ and $L \to \pi R $. For this  purpose it is convenient to define the variable $t\equiv 1/y^2$. Then 

\begin{equation}
L =\! \! R\sqrt{1\! + \! \Lambda^2} \! \! \int_{0}^{1} \! \! dt \frac{t^{1/2}}{\sqrt{1-t}\sqrt{t+\Lambda^2}\sqrt{\Lambda^2\! + \!(1+ \! \Lambda^2\! )t}}
\end{equation}

\begin{equation}
F = \! \! {\Lambda R \over 2\pi \alpha'} \! \!\int_{0}^{1} \! \! \! dt \,  t^{-3/2}\! \!\left [\! \! \frac{\sqrt{t+\Lambda^2}}{\sqrt{1\! - \! t}\sqrt{\! \Lambda^2\! +\!(1\!+\! \Lambda^2\! )t}}\! \! - \!  \!1 \!\! \right ]\!\! -\! {\Lambda R \over \pi \alpha'}
\end{equation}

Now we can evaluate the asymptotic limits . 

\vspace{.4 true cm}

\hspace{1 true cm}{\bf(i)  $L << R$ }

\vspace{.4 true cm}

In this case $\Lambda $ is very large, then we have 

\begin{eqnarray}
L\! \! & \! \approx& \! \! \! \frac{R}{\Lambda}\! \! \int_{0}^{1} \! \! \! dt \, t^{1/2}(1-t^2\!)^{-1/2} \! \! = \! \! \frac{R}{2\Lambda}\! \! \int_{0}^{1} \! \! \!d\rho \, \rho^{-1/4}(1-\rho)^{-1/2} \nonumber \\
& = &\! \! \! \frac{R}{2\Lambda} B(3/4,1/2)
\end{eqnarray}

\noindent where $\rho=t^2$ . The integral for the free energy F is

\begin{equation}
F \approx  {\Lambda R \over 2\pi \alpha'} \int_{0}^{1} dt \,  t^{-3/2}\left [\frac{1}{\sqrt{1-t^2}}-1 \right ]- {\Lambda R \over \pi \alpha'}=  -{a \over L} \label{coulomb}
\end{equation}
where $a = {2C_1^2R^2 \over \pi \alpha'}$ with $C_1 = 1.198$. This result is similar as the one obtained working in the Poincar\'e $AdS$ metric \cite{MaldaPRL}. For small distances $x$ this metric can be thought as the asymptotic limit $r >> R$ of our global $AdS$ metric , which is

\begin{equation}
ds^2 =  {r^2 \over R^2}d t^2 + {R^2 \over r^2} d r^2 + 
{r^2\over R^2} dx ^2\label{metricapoincareads}
\end{equation}

\noindent but note that in this case the coordinate $x$ has to be small. The behavior of this free energy is analogous to that of the phenomenological Cornell potential for a heavy quark anti-quark pair at small distance . 

\vspace{.4 true cm}

\hspace{1 true cm}{\bf(ii)  $L \to  \pi R$ } 

\vspace{.4 true cm}

In this case $\Lambda$ is small then the integral for L is 

\begin{eqnarray}
L &\approx& R \,\int_{0}^{1}dt \, \frac{t^{1/2}}{\sqrt{1-t}(t+\Lambda^2)}
\nonumber \\
&=& \pi \, R \,[ 1 - \frac{\Lambda}{\sqrt{1+\Lambda^2}}] \label{L}
\end{eqnarray}

The integral for the free energy is given by 

\begin{eqnarray}
F &\approx& \! \! {\Lambda R \over 2\pi \alpha'}\! \! \! \int_{0}^{1}\! \! \! \! dt \,  t^{-3/2}\! \! \left [\! \! \frac{1}{\sqrt{1\!-\!t}}\Big(\!1\! - \!\frac{\Lambda^2t}{2(2\Lambda^2\! + \!t)} \!\Big)\! \! - \!1  \! \right] \! \!- \! \!{\Lambda R \over \pi \alpha'}
\nonumber \\
 &=& {\Lambda R \over 4 \pi \alpha'} [ {L \over R} - \pi ]
\end{eqnarray}
where we approximated the term 

\begin{equation}
\frac{1}{\sqrt{\Lambda^2+(1+\Lambda^2)t}}\approx \frac{1}{\sqrt{\Lambda^2+t}}(1-\frac{t\Lambda^2}{2(\Lambda^2+t)})
\end{equation}

From equation (\ref{L}) we have that 

\begin{equation}
\Lambda = -\frac{\frac{L}{ R} - \pi}{\pi\sqrt{1-(1-\frac{L}{\pi R})^2}}
\end{equation}

\noindent so we finally obtain 

\begin{eqnarray}
F &\approx& - \frac{R}{4\pi^2 \alpha'}\frac{({L \over R}-\pi)^2}{\sqrt{1-(1-\frac{L}{\pi R})^2}} \noindent \\
&= &- \frac{R}{4\pi^2 \alpha'}({L \over R}-\pi)^2 \label{energy1}
\end{eqnarray}

This result could be obtained considering the asymptotic behavior $r<<R$ of the metric :

\begin{equation}
ds^2 =  d t^2 + d r^2 + 
{r^2\over R^2} dx ^2\label{metricaminkowski}
\end{equation}
with the quarks in a brane localized at  $r_1 = 2R/\pi$.
 
Note  that when $L$ reaches $L_{\rm max} = \pi R $ the free energy takes its maximal value $F =0$.

\section{Black Hole metric}
The black hole $AdS_4$ metric in global coordinates is defined by

\begin{eqnarray}
ds^2 &\! \! =&\! \! \Big({r^2\over R^2} +1 - {w_3M\over r}\Big)dt^2 + {dr^2 \over \Big({r^2\over R^2} +1 - {w_3M \over r}\Big)} + \nonumber \\
& & r^2\Big(d\theta^2 + \sin^2\theta \, d\varphi^2\Big)
\end{eqnarray}

\noindent where $0\le \theta < \pi $ ,  $-\pi \le \varphi <\pi $ and $0 \le t \le \beta$  . The factor $w_3$ is included so that $ M$ is the mass of the black hole. Also, the spacetime is restricted to the region $r\geq r_{+}$, with $r_{+}$ the largest solution of the equation

\begin{equation}
{r^2 \over R^2} + 1 - {w_3 M \over r} =0\, .
\end{equation}

The region $r=r_+$ is the horizon of the black hole metric . The temperature of the black hole is related to $r_+$ by
\begin{equation}
\frac{1}{T}=\beta = \frac{4\pi R^2 r_+}{3r_+^2 +R^2}
\end{equation}

 Note that there is a minimum temperature $T_{min}$ for the existence of this black hole that corresponds to a critical horizon $r^{c}_{+}=R/\sqrt{3}$. This critical horizon divides two possible black hole solutions : the big black hole (BBH) with an horizon that grows with temperature and the small black hole (SBH) with an opposite behavior. The Hawking-Page criteria states that for temperatures $T>T_{\rm min}$ the big black hole is stable only for  $r_+\ge R$ while the small black hole is always instable \cite{Hawking:1982dh}. 

We again consider $\theta = \pi/2$ and define  $x= R \, \varphi$ so the black hole metric reads 

\begin{equation}
ds^2 \! \! =\! \! \Big({r^2\over R^2} +1 - {w_3M\over r}\Big)dt^2 + {dr^2 \over \Big({r^2\over R^2} +1 - {w_3M \over r}\Big)} + {r^2\over R^2} \, dx^2 \, ,
\end{equation}

\noindent then
\begin{eqnarray}
f^2 &=& g_{tt} g_{x  x} =({r^2\over R^2} +1 - {w_3M\over r}\Big){r^2 \over R^2}\nonumber \\
g^2 &=& g_{tt} g_{r r} = 1
\end{eqnarray}

\begin{figure}
\vspace{.2cm}\centering
\includegraphics[width=5.5cm]{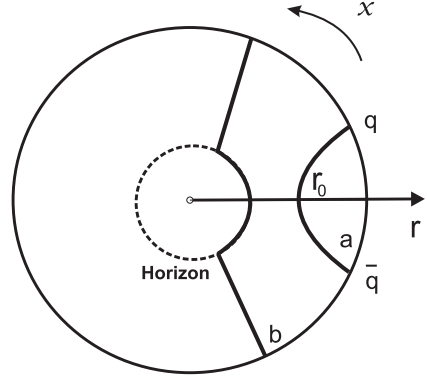}
{\caption{$\! \!$ Curve $a$, correspond a U-type geodesic  and $b$ a Box-type geodesic. }\label{graficocaixote}} 
\end{figure} 

Now, we put a quark at  $x = {L \over 2}$ and an anti-quark at $x =  -{L \over 2}$. In this space we have two possible solutions for the static string : the first one is a U-type solution  similar to the one obtained before for the $AdS$ where the string reaches a minimum for the radial coordinate $r_0$ , the second solution is a string reaching the horizon (see Fig \ref{graficocaixote}). Depending on the value of $L$ one of those solutions  will have the  minimum free energy. The distance and free energy for the quark anti-quark pair are in this case

\begin{widetext}

\begin{eqnarray}
L & = & 2R \sqrt{1 + \Lambda^2 - {\gamma \over \Lambda}}\int_0^1 {dt \, t^2 \over \sqrt{\Lambda^2 + t^2 -{\gamma \over \Lambda}t^3} \sqrt{1 -t}\Big[\Big(1+t\Big)\Big( (1+ \Lambda^2)t^2 + \Lambda^2\Big) - {\gamma \over \Lambda}t^3\Big]^{1/2}}
\\
& & \nonumber\\
& & \nonumber \\
F &=& \frac{R\Lambda}{\pi \alpha '} \int_0^{1} dt \, t^{-2}\Big[ \frac{\sqrt{\Lambda^2+t^2-\frac{\gamma}{\Lambda }t^3}}{\sqrt{1-t}[(1+t)(\Lambda^2 +(1+\Lambda^2)t^2)-\frac{\gamma}{\Lambda }t^3]^{1/2}}-1 \Big] - {R\over \pi \alpha' } \Big(\Lambda -\Lambda_+\Big) \, \label{eqenergyBH}. 
\end{eqnarray}

where $t \equiv {r_0 \over r}$, $\Lambda = {r_0 \over R}$ and  we defined

\end{widetext}

\begin{figure}
\centering
\includegraphics[width=9cm]{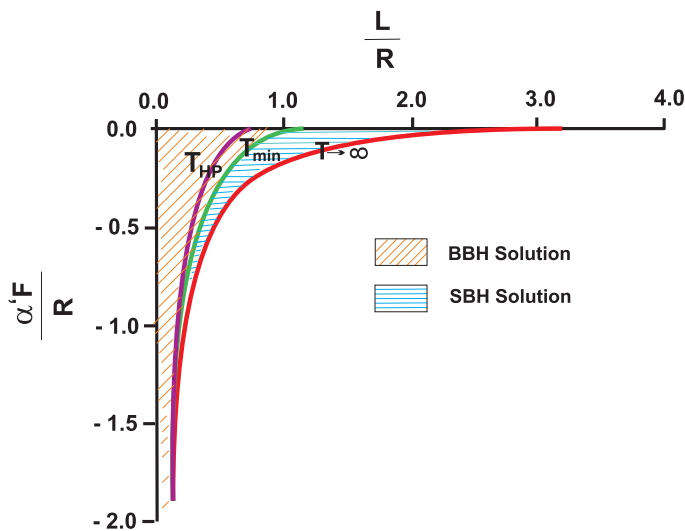}
{\caption{$F\alpha ' \over R $\, vs \,  $L\over R$ \, for black hole $AdS_4$ at different temperature values. \label{energiaBH}}} 
\end{figure}

\begin{equation}
\gamma \equiv {\omega_3 M \over R} = \Lambda_+\Big(1+\Lambda_+^2 \Big ) 
\end{equation}

\noindent with $\Lambda_+ \equiv \frac{r_+}{R}$. Note that $\Lambda \ge \Lambda_+$ because $r_0 \ge r_+$ . We also used the mass regulator


\begin{equation}
2m_q = {1 \over \pi \alpha'} \int_{r_+}^{\infty} dr \, g(r)
\end{equation}

From eq (\ref{eqenergyBH})  it is not difficult  to show that the Box-type solution has always zero free energy. The free energy of the U-type solution is negative for small $L$ and exhibits the same coulombian behavior found in eq. (\ref{coulomb}) .  When we increase the distance the free energy increases and reaches the zero value for some length $L_c$ that depends on  $\Lambda_+$. When $L>L_c$ the free energy of the U-type solution becomes positive and is inestable compared with the Box-type solution (which has zero free energy for every $L$). So we find a transition for the free energy corresponding to a transition between the U-type and the Box-type solutions being $L_c$ a screening length for which the free energy reaches the zero value. We show in Fig. \ref{energiaBH} the free energy as a function of the distance for the two black hole solutions at different values of the temperature. Note  that the big black hole solution has a screening length that is always lower than the maximal distance $L_{\rm max}=R \pi $ while the small black hole has a confining limit corresponding to $L_c\to L_{\rm max}$.

\begin{figure}
\centering
\includegraphics[width=7cm]{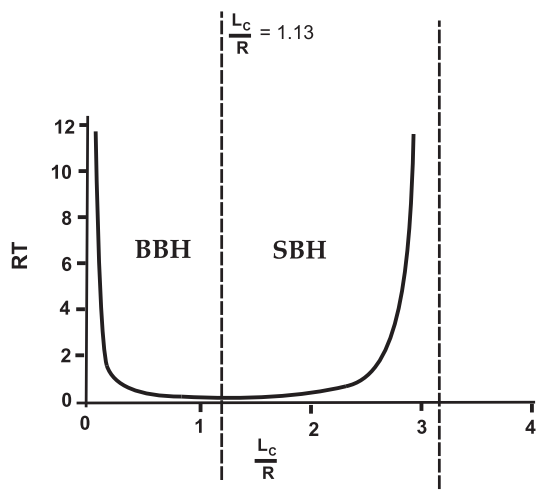}
{\caption{$T$\, vs \, $L^{\rm crit}$  \, for black hole $AdS_4$ . \label{tlcrit}}} 
\end{figure}

\begin{figure}
\centering
\includegraphics[width=8.5cm]{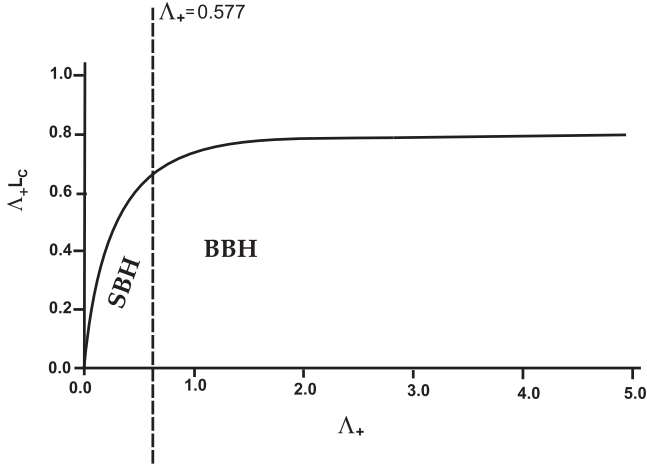}
{\caption{$\Lambda_+ L_c$  \, vs \, $\Lambda_+$  \, for black hole $AdS_4$ . \label{lcritlambdamais}}} 
\end{figure}

This limit is achieved when $T \to \infty$   that corresponds to $\Lambda_+\to 0$ when  the small black hole $AdS_4$ becomes $AdS_4$ (which is confining). In Fig. \ref{tlcrit} we show the temperature as a function of the screening length. It is interesting to obtain the screening length dependence on $\Lambda_+$ for the big black hole. From Fig. \ref{lcritlambdamais} we see that for large $\Lambda_+$ the product $\Lambda_+ L^{\rm crit}$ is constant and approximately equal to $0.79$. This behavior was also obtained for the Poincar\'e black hole \cite{BoschiFilho:2006pe}.

\newpage

\section{What about $\bm    S^{n-1}\times S^1$ ?}

The quark anti-quark potential in $S^{n-1}\times S^1$ can be calculated in a similar way as the $S^2\times S^1$ case. At low temperatures we have to deal with the global $AdS_{n+1}$ space. The metric of this space  for $n\ge 4$ is given by

\begin{eqnarray}
ds^2 \! \! &=& \! \!\Big(\! 1 \!+ \!{r^2 \over R^2} \! \Big) d t^2 \!+\! {d r^2 \over \Big( 1 + {r^2 \over R^2} \Big)}\! +\! r^2\Big( d \theta_1^2 +\prod_{ i=1}^{n-2}\sin^2\theta_i \,d\varphi ^2\Big) \nonumber \\
&+ &  \sum_{m=2}^{n-2} \prod_{ i=1}^{m-1}r^2 \sin^2\theta_i \,d\theta_m ^2 \label{metricaadsndim}
\end{eqnarray}

\noindent where $0\le \theta_1 < \pi$ \,  $0 \le \theta_m < \pi $ ,  $-\pi \le \varphi <\pi $ , $0<r<\infty$ and $0 \le t \le \beta$.
If we choose $\theta_1=\theta_m=\pi/2$ and define $x=R\varphi$ we arrive at the same metric of eq.(\ref{metricaadsx}) so we find the same free energy found before working in $AdS_4$ (Fig \ref{Fig5}).

\newpage
At high temperatures, we use the black hole $AdS_{n+1}$ ($n\ge4$) with metric 

\begin{eqnarray}
ds^2 \! \!\! \! &=& \! \! \!\Big(\! 1 \!+ \!{r^2 \over R^2}- {w_n M\over r^{n-2}} \! \Big) d t^2 \!+\! {d r^2 \over \Big( 1 + {r^2 \over R^2} -{w_n M\over r^{n-2}}\Big)}\! + r^2 d \theta_1^2\! \!\nonumber \\ \! \!&\! \! + \! \!& \! r^2 \prod_{ i=1}^{n-2}\! \sin^2\theta_i \,d\varphi ^2\! 
 +  \sum_{m=2}^{n-2}\! \prod_{ i=1}^{m-1}\! r^2 \! \sin^2\! \theta_i \,d\theta_m ^2 \label{metricabhadsndim}
\end{eqnarray}
\noindent where $r \ge r_+$. The temperature is now related to $r_+$ by 
\begin{equation}
\frac{1}{T}=\beta = \frac{4\pi R^2 r_+}{n r_+^2 +(n-2)R^2}
\end{equation}

Now the critical horizon is $r^{c}_{+}=R\sqrt{\frac{n-2}{n}}$ , the Hawking-Page horizon is always   $r_+= R$ which is always above $ r^{c}_{+}$ so the big black hole ($r_+ \ge  R$) is stable while the small black hole is always instable \cite{Hawking:1982dh}. Choosing again $\theta_1=\theta_m=\pi/2$ and defining $x=R\varphi$ we obtain
\begin{equation}
ds^2 \! \! =\! \! \Big({r^2\over R^2} +1 - {w_n M\over r^{n-2}}\Big)dt^2 + {dr^2 \over \Big({r^2\over R^2} +1 - {w_n M \over r^{n-2}}\Big)} + {r^2\over R^2} \, dx^2 \, ,
\end{equation}

\noindent so we find in this case 
\begin{equation}
\begin{array}{ll}
f^2 = g_{tt} g_{x  x} =({r^2\over R^2} +1 - {w_n M\over r^{n-2}}\Big){r^2 \over R^2}\\
g^2 = g_{tt} g_{r r} = 1
\end{array}
\end{equation}

\noindent leading to 

\begin{widetext}

\begin{eqnarray}
\! \! \! \! \! L \! \! \! &=&  \! \! \!2R \! \! \int_0^1 \! \! \! \! \! \!{dt \, t^2 \sqrt{1 + \Lambda^2 - {\gamma \over \Lambda^{n-2}}} \over \sqrt{\Lambda^2 + t^2 -{\gamma \over \Lambda^{n-2}}t^n} \Big[(1-t^2)((1+ \Lambda^2)t^2 + \Lambda^2)+\frac{\gamma}{\Lambda^{n-2}}(t^4-t^n)\Big]^{1/2}}
 \nonumber\\
& & \nonumber \\
& & \nonumber \\
\! \! \! \! \! F \! \! \!  &=& \! \! \!\frac{R \!\Lambda}{\pi  \alpha '} \! \!  \! \! \int_0^{1}\! \! \! \! \! dt \, t^{^{\! -2}} \!\left[\! \!\frac{\sqrt{\! \Lambda^2\! + \! t^2 \! - \! \frac{\gamma}{\Lambda^{n-2} }t^n
}}{[(1-t^2)( (1\! + \Lambda^2 )t^2+\Lambda^2)+\frac{\gamma}{\Lambda^{n-2}}(t^4-t^n)]^{1/2}}\! -\! 1\! \right] \! \! -\! \frac{R }{\pi  \alpha '}\Big(\Lambda - \Lambda_+ \Big) \!\nonumber \\ & & \label{eqenergyBH2}
\end{eqnarray}
\noindent where 
\begin{equation}
\gamma \equiv {\omega_n M \over R^{n-2}} = \Lambda_+^{n-2}\Big(1+\Lambda_+^2 \Big )  \, .
\end{equation}

\end{widetext}

\begin{figure}
\centering
\includegraphics[width=8 cm]{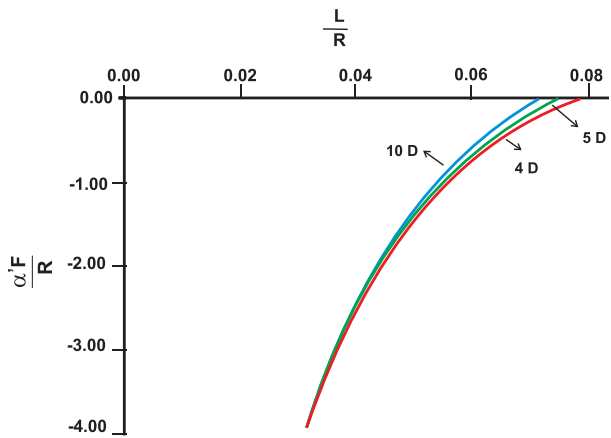}
{\caption{$ \! \!$  $\bm{\alpha' F\over R}\!$ \, vs \, $ \! \bm {L\over R}$  \, for black hole in $AdS_4$\,,\,$AdS_5$ and $AdS_{10}$\\ ($\bm \Lambda_+ \!= \!10 $\,). \label{energiaBH2}}} 
\end{figure} 

These integrals are slightly different from the integrals found before but the result for the free energy is very similar. We find (again) a transition between a U-type and a Box-type solution and a screening length $L_c$  so that the free energy is negative below $L_c$ and zero above. We compare in Fig \ref{energiaBH2} the potential energies as a function of the distance for $n=3,4$ and $n=9$ at a fixed horizon position $\Lambda_+=10$. We see that the energies coincide at small distances and there is a little diminution of the screening length when the dimension grows up. Then the screening length will never reach the maximum distance $L_{\rm max}=R \pi$ in the big black hole. This means that the corresponding gauge theory in $S^{n-1}\times S^1$ is deconfining at high temperatures for any $n \ge 3$. For the special case $n=4$ which corresponds to SYM theory in $S^3\times S^1$ our results agree with those obtained before by Landsteiner and Lopez  \cite{Landsteiner:1999up} where this screening length was defined. 

\section{Physical discussion}

We calculated the free energy of a quark anti-quark pair in a compact space $S^{n-1}\times S^1$ with the period of $S^1$ being the inverse of the temperature. For the global $AdS$ we find a quark anti-quark free energy that has a coulomb-like behavior and that goes to zero only at the maximum distance $L_{\rm max}$ in the circle of Fig \ref{Fig2}. So if the quark is in a fixed position the only way to obtain a zero free energy is to put the anti-quark at a distance $L_{\rm max}=\pi R$. This means that the anti-quark does not have the freedom to move in the $x$ coordinate without feeling the potential so this corresponds to a confinement scenario. This scenario is in accordance with the one obtained in \cite{Witten:1998zw}. At high temperatures the static string in the black hole metric leads to a coulomb-like free energy that goes to zero at a screening length $L_c$. We calculate the dependence of $L_c$ on the horizon position and temperature and show that it only reaches $L_{\rm max}$ for the small black hole solution which is instable by the Hawking-Page criteria. For the (stable) big black hole we found that $L_c$ is always lower than  $L_{\rm max}$ so considering  a quark in a fixed position the anti-quark has some freedom to move in the $x$ coordinate with corresponds to a deconfinement scenario.  We conclude that the transition of the quark anti-quark free energy in $S^{n-1}\times S^1$ when one goes from $AdS$ to black hole $AdS$ is indeed a confinement/deconfinement transition when included the Hawking-Page criteria of stability. This confirms the interpretation of the confinement/deconfinement transition in $S^{n-1}\times S^1$ as the dual of the Hawking-Page transition supported by the entropy jump of the spaces.   

\noindent {\bf Acknowledgments}: We would like to thank Nelson Braga and Henrique Boschi  
for important suggestions and comments. We also thank the CBPF for the hospitality . The authors are partially supported by CNPq and CLAF.

\end{document}